\documentclass{article}
\usepackage{dcase2024,amsmath,graphicx,url,times,booktabs, tabularx}

\usepackage{multirow}
\usepackage{array}
\usepackage{makecell}

\title{A sound description: exploring prompt templates and class descriptions to enhance zero-shot audio classification}

\name{Michel Olvera,
        Paraskevas Stamatiadis, 
      Slim Essid
      }
\address{ LTCI, Télécom Paris, Institut Polytechnique de Paris, France  \\         
        \{olvera, paraskevas.stamatiadis, slim.essid\}@telecom-paris.fr\\
 }

\newcommand\blfootnote[1]{%
  \begingroup
  \renewcommand\thefootnote{}\footnote{#1}%
  \addtocounter{footnote}{-1}%
  \endgroup
}

\begin{document}

\ninept
\maketitle

\begin{sloppy}

\begin{abstract}
Audio-text models trained via contrastive learning offer a practical approach to perform audio classification through natural language prompts, such as ``this is a sound of" followed by category names. In this work, we explore alternative prompt templates for zero-shot audio classification, demonstrating the existence of higher-performing options. First, we find that the formatting of the prompts significantly affects performance so that simply prompting the models with properly formatted class labels performs competitively with optimized prompt templates and even prompt ensembling. Moreover, we look into complementing class labels by audio-centric descriptions. By leveraging large language models, we generate textual descriptions that prioritize acoustic features of sound events to disambiguate between classes, without extensive prompt engineering. We show that prompting with class descriptions leads to state-of-the-art results in zero-shot audio classification across major ambient sound datasets. Remarkably, this method requires no additional training and remains fully zero-shot.

\end{abstract}


\vspace{-0.5em}
\section{Introduction}
\label{sec:intro}
Multimodal contrastive pretraining has been used to train multimodal representation models on large amounts of paired data. This approach leverages contrastive learning to align representations across different modalities, promoting a shared embedding space that improves semantic understanding across modalities. Examples include Contrastive Language-Image Pretraining (CLIP) \cite{radford2021learning}, which aligns visual and textual representations, and the more recent Contrastive Language-Audio Pretraining (CLAP), which extends these principles to align audio and textual representations \cite{guzhov2022audioclip, elizalde2024natural, wu2023large, manco2022contrastive}.

Following pretraining, CLAP exhibits a well-structured feature space, yielding robust, general-purpose representations well-suited for downstream training. Moreover, it also demonstrates exceptional transferability as evidenced by its impressive zero-shot performance across classification, captioning, retrieval, and generation tasks  \cite{elizalde2024natural, mei2023wavcaps, liu2023audioldm}.

Extensive research on CLIP has revealed that classification scores are significantly influenced by alterations in prompt formulation and language nuances. For instance, varying the description of a concept, using synonyms, or modifying the grammatical structure or wording, substantially affects performance outcomes \cite{yuksekgonul2023and, anperceptionclip, salinas2024butterfly}. Besides, prompts offering more context or specificity tend to yield more accurate results \cite{pratt2023does, roth2023waffling, mirza2024lafter}.\blfootnote{This work was supported by the Audible project, funded by French BPI.}

Similarly, CLAP inherits sensitivity to prompting from its contrastive pretraining approach. Yet, the systematic exploration of prompt robustness in CLAP remains limited, despite few works highlighting the sensitivity of classification to prompt variations \cite{elizalde2023clap,kushwahamultimodal}. These works, primarily conducted on the ESC50 dataset and limited to up to five prompt templates, shed initial light on these variations. However, robustness to prompt changes is likely to vary across different datasets. Addressing this gap, recent efforts have explored alternative approaches, such as prompt tuning strategies and lightweight adapters, to mitigate the reliance on manually engineered prompts \cite{li2024audio, deshmukh2024domain} with an explicit focus on adapting CLAP to downstream tasks or new domains.

In this work, we propose a tuning-free approach that prompts CLAP models with descriptions of class labels to enhance zero-shot audio classification. While using keywords such as “audio,” “hear,” and “sound” in prompt templates primes the text encoder to focus on audio-related concepts, we hypothesize that enriching prompts with explicit class descriptions can further enhance the model’s ability to clarify the meaning of class labels, particularly in scenarios where labels are ambiguous.
Ambiguity stems from both the textual and audio aspects of the data. Textual ambiguity arises from homonyms, where words possess multiple meanings, and from the lack of contextual clues (\textit{e.g.}, "bat" as both an animal and sports equipment). On the audio side, ambiguity arises from acoustically similar sound categories, such as distinguishing between bird vocalizations (\textit{e.g.}, raven vs. crow calls) and musical instruments (\textit{e.g.}, violin vs. viola). Thus, detailed prompts may clarify sounds heavily reliant on context, and help disambiguate acoustically similar sounds.
Such descriptions can also disambiguate abstract sounds such as ``white noise" and compensate for knowledge gaps or limited exposure to certain terms. For instance, clarifying “Geiger counter”, as “a detection device that clicks or beeps when detecting radiation” could improve correlations of audio and text features.

\begin{table*}[t!]
\centering
\scriptsize
\resizebox{\textwidth}{!}{
\begin{tabularx}{\linewidth}{p{1cm}p{4cm}p{6cm}p{5cm}}
\toprule
Class & Base & Context & Ontology \\
\midrule
Mandolin & A stringed musical instrument, played with a plectrum, characterized by its small size, high-pitched sound, and distinctive twang. & A stringed musical instrument with a distinctive, twangy sound, often associated with folk or bluegrass music. Typically played by plucking or strumming the strings, producing a bright, melodic tone. & A stringed musical instrument with a distinctive, twangy sound, often used in folk and pop music. \\
Rail transport & The sound of trains moving on rails, characterized by the clacking of wheels and the rumbling of engines. & The sound of trains moving along rails, characterized by a steady, rhythmic clacking or clicking noise. Often heard in urban or rural areas with rail infrastructure. & The rumbling and clanking sounds produced by trains moving on rails, characterized by their speed and intensity, classified under transportation-related sounds. \\
Toot & A short, high-pitched sound produced by blowing air through a small opening, often used as a signal or warning. & A short, sharp sound, typically produced by blowing air through a small opening, such as a whistle or a musical instrument. & A short, high-pitched sound produced by a whistle or other musical instrument, often used as a signal or warning. \\
Stream & A continuous flow of water or other liquid, often characterized by its sound as it flows over rocks or other obstacles. & A continuous flow of water, often heard in natural environments like rivers, lakes, or waterfalls, characterized by the sound of water flowing over rocks or other surfaces. & A continuous flow of sound, often characterized by its rhythmic patterns and timbre, belonging to the category of natural environmental sounds. \\
\bottomrule
\end{tabularx}
}
\caption{Example descriptions of randomly sampled class labels from the datasets considered in this work, generated with Mistral-7B \cite{jiang2023mistral}.}
\label{tab:example_definition}
\end{table*}

To validate our hypothesis, we leverage Large Language Models (LLMs) for their knowledge of sound semantics. Specifically, we used Mistral\footnote{https://huggingface.co/mistralai/Mistral-7B-Instruct-v0.1} to describe the acoustic properties of class labels. Our study demonstrates that using audio-centric descriptions of class labels as prompts helps CLAP better ground acoustic features with semantic descriptions, significantly boosting zero-shot classification scores across major environmental sound datasets. Remarkably, our method even outperforms learnable prompt strategies, all without the need for additional training, while remaining entirely zero-shot.





\vspace{-1em}
\section{Methodology}
\label{sec:methodology}
We first describe the zero-shot audio classification task, then our adaptive class selection strategy and finally we motivate our LLM-generated  class descriptions. 

\subsection{Zero-shot audio classification}
Given a set of target categories \( C \) and a query audio sample \( a \), the zero-shot audio classification protocol in CLAP defines the classification problem as a nearest neighbor retrieval task. The predicted category \(\hat{c}\) is determined as follows:
\begin{equation}\label{eq:setup_class}
    \hat{c} = \arg\max_{c \in C} \text{sim}(\phi_{\text{A}}(a), \phi_{\text{T}}(c)),
\end{equation}  
where \( C \) represents the set of class labels, \( a \) denotes the input audio, and \(\phi_{\text{A}}\) and \(\phi_{\text{T}}\) are the audio and text encoders, respectively. The function \(\text{sim}(\cdot, \cdot)\) corresponds to the similarity metric, typically the cosine similarity.

To enhance zero-shot audio classification, we propose using both class labels and their descriptions to resolve ambiguities. Given a set of target categories \( C \), definitions \( D \), the predicted category $\tilde{c}$ is determined by:

\begin{equation}\label{eq:setup_descrtiption}
    \tilde{c} = \arg\max_{c \in C} \text{sim}(\phi_{\text{A}}(a), \phi_{\text{T}}(c + d_c))),
\end{equation}
where $d_c \in D$ is the description corresponding to class $c$, and the $+$ operator denotes the textual combination of the class label $c$ and its description $d_c$.


\subsection{Adaptive class description selection}
\label{subsec:adaptive_selection}
We devise an adaptive strategy that incorporates descriptions selectively for classes potentially ambiguous to the text encoder.
Let $\text{P}_\text{class-only}$ and $\text{P}_\text{class-description}$ represent the classification performance for class $c$ using setups involving classes only or classes with descriptions as in Equations \eqref{eq:setup_class} and \eqref{eq:setup_descrtiption}, respectively. We decide for class $c$ which setup to apply through the decision function $\text{M}(c)$: 

\begin{equation} \label{eq:mapping_xval}
    \text{M}(c) = 
    \begin{cases} 
        \hat{c} & \text{if } \text{P}_\text{class-only} \geq \text{P}_\text{class-description} \\
        \tilde{c} & \text{if } \text{P}_\text{class-description} > \text{P}_\text{class-only}.
    \end{cases}
\end{equation}

The function $\text{M}(c)$ decides whether a class should include a description based on cross-validation of results.





\subsection{Generation of audio-centric descriptions with LLMs}
Given audio event class labels, we propose to use Large Language Models (LLMs) to generate audio-centric descriptions for them automatically, as manual collection of descriptions entails a labor-intensive endeavor. LLMs, trained on vast text data, have a deep understanding of language, which we exploit for their knowledge of sound semantics. Our method, adapted from \cite{oncescu2024sound}, involves three steps. First, we provide a general description of the task. Second, we combine these instructions with in-context demonstrations, including a few paired label-description examples. Finally, we provide the LLM with the class labels, heuristic constraints, and specific output format details to generate audio-centric descriptions.

Using this method, we generated three types of descriptions: 
\textit{base descriptions}, \textit{context-aware descriptions}, and \textit{ontology-aware descriptions}. All are audio-centric. \textit{Base descriptions} reflect the acoustic properties and characteristic sounds of the class labels. \textit{Context-aware} descriptions add details about the typical locations and circumstances of encountering the sounds, including the physical environment, associated objects, and the function of the sound within its context. \textit{Ontology-aware} descriptions capture the acoustic properties and characteristic sounds of each class label while also considering their relationships with coarse high-level concepts. Table \ref{tab:example_definition} provides a few examples of the generated descriptions. The complete list of class descriptions and the prompts used to generate them are available on our companion website.\footnote{https://github.com/tpt-adasp/a-sound-description}

\section{Experimental Setup}
\label{sec:experimental_setup}
We detail our experimental approach, including model and dataset selection, evaluation metrics, and experiments to explore different prompt strategies and their impact on classification.

\subsection{Models}
We adopt two state-of-the-art audio-text models pre-trained via contrastive learning, namely LAION-CLAP (LA) and Microsoft CLAP 2023 (MS). The former utilizes RoBERTa \cite{liu2019roberta} as its text encoder, while the latter leverages GPT-2 \cite{radford2019language}. Both models rely on HTS-AT \cite{chen2022hts} as their audio encoder.

\subsection{Datasets and evaluation metrics}
\textbf{Downstream datasets}. We select six major environmental sound datasets tailored for either single-class or multi-label classification. These include: \textbf{ESC50} \cite{piczak2015dataset}, which contains 50 environmental sound classes with 2k labeled samples of 5 seconds each; \textbf{US8K} \cite{salamon2014dataset}, comprising 10 urban sound classes and 8k labeled sound excerpts of 4 seconds each; \textbf{TUT2017} \cite{mesaros2019sound}, consisting of 15 acoustic scenes classes and 52k files of 10 seconds each; \textbf{FSD50K} \cite{fonseca2021fsd50k}, featuring 51K audio clips of variable length (from 0.3 to 30 seconds each) curated from Freesound and comprising 200 classes; \textbf{AudioSet} \cite{gemmeke2017audio}, a large-scale dataset encompassing 527 classes, with over 2 million human-labeled sound clips of 10 seconds from YouTube videos; and \textbf{DCASE17-T4} \cite{mesaros2019sound}, a subset of AudioSet focused on 17 classes related to warning and vehicle sounds, containing 30k audio clips of 10 seconds each.
\\\\
\textbf{Evaluation setup and metrics}. In our evaluation we consider all available splits (train/val/test) or folds, except for AudioSet, where only the test set was used. Note that some datasets do not allow for a fully zero-shot approach, as some audio files used in the evaluation were part of the pretraining data of the considered  frozen CLAP models (\textit{e.g.}, AudioSet and FSD50K). We believe that it is still interesting to analyse the corresponding results, bearing this fact in mind during the discussion. We use accuracy as the metric for single-class classification datasets (ESC50, US8K and TUT2017) and mean Average Precision (mAP) for multi-label classification datasets (FSD50K, AudioSet and DCASE17-T4). For experiments involving class-specific descriptions, a 5-fold cross-validation setting is employed. These folds were constructed on the data considered for evaluation \textit{i.e.,} all splits/folds for all datasets, except for AudioSet where the test set is used. In this approach, training folds are used to derive the mapping $\text{M}$ from Equation \eqref{eq:mapping_xval}, while test folds are used to assess its generalization. Directly evaluating the mapping without cross-validation would yield overly optimistic results due to overfitting.

\vspace{-0.5em}
\subsection{Zero-shot audio classification experiments}
\textbf{Prompting with class labels only}
We explore zero-shot audio classification using prompts with sanitized class labels (\textit{i.e.}, replacing underscores in original labels with spaces, \textit{e.g.}, \textit{dog\_barking} becomes \textit{dog barking}). This is motivated by the fact that in our early experiments we observed that this strategy performs competitively compared to prompting with ``This is a sound of'', which has been preferred in the literature \cite{elizalde2023clap, wu2023large}. Here, we systematically study the impact of using only class labels as prompts on classification performance. We examine four different formats to construct the start and end of a prompt: uppercase with a period (\textit{e.g., Dog barking.}), uppercase without a period (\textit{e.g., Dog barking}), lowercase with a period (\textit{e.g., dog barking.}), and lowercase without a period (\textit{e.g., dog barking}). The format yielding the highest performance for each model, termed as CLS, was selected as a reference for subsequent experiments involving class descriptions.
\\ \\
\textbf{Prompting with templates}. Inspired from CLIP \cite{radford2021learning}, we explore a set of prompt templates as plausible alternatives to ``This is a sound of", all tailored for the zero-shot audio classification task. We curated a set of 33 distinct prompts, drawing some from prior studies \cite{elizalde2023clap, wu2023large, kushwahamultimodal}. Our objective is to systematically evaluate the performance of these alternative prompts and their ensemble across multiple datasets. Each prompt follows the format \textit{Template + class label}, \textit{e.g.}, ``A sound clip of dog barking.". We thus analyse the performance of three prompt configurations:
PT$_{\text{Baseline}}$: The baseline prompt template ``This is a sound of".
PT$_{\text{Best}}$: The most effective prompt template identified among the 33 manually crafted alternatives.
PT$_{\text{Ensemble}}$: Ensembling text embeddings from all considered prompt templates.
Each prompt template begins with an uppercase letter and concludes with a period.
\\\\
\textbf{Prompting with class-specific descriptions}.
We investigate the impact of combining class labels and their descriptions generated by LLMs. The experimental setups include:
CLS: Class label only.
CD$_{\text{Base}}$: Audio-centric definitions generated by Mistral.
CD$_{\text{Context}}$\footnote{We did not consider context-aware descriptions for TUT2017 because these were very similar to base descriptions. Unlike other datasets, TUT2017 comprises labels that refer to acoustic scenes. This explains the similarity, as both type of descriptions indicate context.}: Context-aware descriptions.
CD$_{\text{Ontology}}$: Ontological information related to the class label.
CD$_{\text{Dictionary}}$: Definitions (non audio-centric) sourced from the Cambridge Dictionary of English.\footnote{When definitions where not available in the Cambridge Dictionary, definitions were sourced from WordNet, Wikipedia, and FreeBase.}

\vspace{-0.5em}
\section{Results and discussion}
\label{se:results}
In this section, we present and discuss the outcomes of our experiments, shedding light on the impact of various prompting strategies and the role of class descriptions in classification performance.
\subsection{Sensitivity to prompt format} 
In Table \ref{tab:results_sensitivity}, we report the average classification results across all evaluation datasets to examine the sensitivity of zero-shot classification performance to subtle variations in the input prompt format. We see surprising differences in performance due to minor alterations such as capitalization and punctuation, consistent with findings in \cite{kushwahamultimodal}. A recent work on LLM behavior confirm that these seemingly minor changes in prompt format influence the model’s internal representations, leading to distinct transformations within the embedding space that alter the output probability distribution in ways that affect classification performance \cite{sclarquantifying}.  We observe that, for both models, prompt variations in punctuation, irrespective of capitalization, significantly affect performance more than variations in capitalization without punctuation. Notably, the performance gap between the most and least effective formats was 5.46\% for model LA and 8\% for model MS, pointing out how critical it is to select an optimal format to maximize classification scores. Consequently, subsequent experiments adopted the best-performing format for each model. 

\renewcommand{\arraystretch}{1.1}
\begin{table}[h]
\centering
\begin{tabular}{ccc}
\toprule
\multirow{2}{*}{Prompt format}  & \multicolumn{2}{c}{Model} \\ \cline{2-3}
                        & LA & MS \\ \hline
class label (\textit{e.g., dog barking})                &  0.5059        &  0.5256       \\ 
class label. (\textit{e.g., dog barking.})                &  0.5524        &  $\mathbf{0.5735}$        \\ 
Class label (\textit{e.g., Dog barking})                &  0.5110        &  0.49344      \\
Class label. (\textit{e.g., Dog barking.})                &  $\mathbf{0.5605}$        &  0.5395      \\ 
\bottomrule
\end{tabular}
\caption{Average model performance scores across all datasets for different input prompt formats.}
\label{tab:results_sensitivity}
\end{table}

\renewcommand{\arraystretch}{1.2}
\begin{table*}[ht]
\centering
\resizebox{\textwidth}{!}{
\begin{tabular}{l|cc|cc|cc|cc|cc|cc|cc}
\toprule
\multirow{2}{*}{Method} & \multicolumn{2}{c|}{ESC50} & \multicolumn{2}{c|}{US8K} & \multicolumn{2}{c|}{TUT2017} & \multicolumn{2}{c|}{DCASE17-T4}  & \multicolumn{2}{c|}{FSD50K} & \multicolumn{2}{c}{AudioSet} & \multicolumn{2}{c}{Average}\\ 
                         & LA & MS & LA & MS & LA & MS & LA & MS & LA & MS & LA & MS & LA & MS \\ \hline
CLS                      & 0.9280    & 0.9280    & 0.7980     & 0.8737    & 0.4242    & 0.5717    & 0.4443    & 0.3772    & 0.5409   & 0.5137    & 0.2277    & 0.1764   & 0.5605    & 0.5735       \\ 
\hline
PT$_{\text{Baseline}}$                       &  0.915   & 0.893    & 0.7747    & 0.7855   & 0.4890   & 0.4547    & 0.4670  & 0.4674  & 0.5308    & 0.5052    &  0.2269    & 0.2708    & 0.5672   & 0.5628            \\ 
PT$_{\text{Best}}$                           & 0.9415   & 0.9585   & 0.8133   & 0.8624    & 0.5041   & 0.6192    & 0.5220  & 0.4583  & 0.5765    & 0.5372    & 0.2855     & 0.2708   & 0.6071   & 0.6176           \\ 
\hline
PT$_{\text{Ensemble}}$                      & 0.9295    & 0.95     & 0.7893   & 0.8506    & 0.4944    & 0.6111    & 0.4851   & 0.4075    & 0.5744    & 0.5424    & 0.2560    & 0.2063    & 0.5881    & 0.5946      \\ \hline
\multicolumn{15}{c}{Adaptive class description selection (mean scores across five folds)} \\ \hline
CD$_{\text{Dictionary}}$      & 0.9535 & 0.9205            & 0.8632    & 0.8891            & 0.5770     & 0.5630           & 0.4704            & 0.3776    & 0.5623            & 0.4972    & 0.2727            & 0.1924    & 0.6165            & 0.5733        \\ 
CD$_{\text{Base}}$       & 0.9480 & 0.9505            & 0.8336    & 0.8926            & 0.5790     & \textbf{0.6219}  & 0.4705            & 0.3911    & 0.5654            & 0.5039    & 0.2803            & 0.1963    & 0.6128            & 0.5927        \\ 
CD$_{\text{Context}}$    & 0.9455 & 0.9595            & 0.8597    & 0.8782            & -          & -                & \textbf{0.4742}   & 0.3801    & \textbf{0.5720}   & 0.5128    & \textbf{0.2891}   & 0.2022    & \textbf{0.6281}   & 0.5865        \\ 
CD$_{\text{Ontology}}$       & 0.9495 & \textbf{0.9635}   & 0.8480    & \textbf{0.9017}   & 0.5030     & 0.5670           & 0.4589            & 0.3748    & 0.5676            & 0.5074    & 0.2830            & 0.1998    & 0.6017            & 0.5857        \\ 
CD$_{\text{All}}$        & 0.9491 & 0.9485            &  0.8511   & 0.8904            & 0.5530     & 0.5840           & 0.4685            & 0.3809    & 0.5668            & 0.5053    & 0.2813            & 0.1976    & 0.6142            & 0.5845        \\ \hline\hline

SOTA      & \multicolumn{2}{c|}{0.96 \cite{kushwahamultimodal}}   &  \multicolumn{2}{c|}{0.8526 \cite{deshmukh2024domain}}     &     \multicolumn{2}{c|}{0.5438 \cite{deshmukh2024domain}}     & \multicolumn{2}{c|}{-} &  \multicolumn{2}{c|}{0.52 \cite{kushwahamultimodal}} & \multicolumn{2}{c|}{0.102 \cite{deshmukh2024domain}}      &    \multicolumn{2}{c}{-}        \\ 
\bottomrule
\end{tabular}
}
\caption{Zero-shot classification scores across 6 downstream tasks. Evaluation metrics: Accuracy for ESC50, US8K and TUT2017; mean Average Precision (mAP) for DCASE17-T4, FSD50K and AudioSet.  }
\label{tab:results_extensive}
\end{table*}
\vspace{-1.5em}
\subsection{Comparison of prompting strategies}
In Table \ref{tab:results_extensive}, top-panel, we show results that assess the impact on classification performance when prompting CLAP models using only the class label and various prompt templates and an ensemble of these prompts. Our findings reveal that  using the class label alone (CLS) often yields superior performance compared to the prompt template "This is a sound of" (PT$_{\text{Baseline}}$). Specifically, CLS demonstrates better results than PT$_{\text{Baseline}}$ on the majority of datasets,  with model MS showing an absolute improvement of 1.07\%. However, for model LA, CLS showed a slight underperformance of 0.67\%, largely due to lower scores on the TUT2017 and DCASE17-T4 datasets.

We report the best-performing prompt template, PT$_{\text{Best}}$, among those considered as plausible alternatives to PT$_{\text{Baseline}}$ for each dataset. On average, PT$_{\text{Best}}$ outperformed PT$_{\text{Baseline}}$, with an absolute improvement of 3.99\% and 5.48\% for LA and MS, respectively. The relevance of this result brings to light the existence of better manually crafted prompt templates than \textit{This is a sound of}. Table \ref{tab:best_prompt_templates} lists the best-performing prompt template for each evaluation dataset. Interestingly, the absence of a ``universal'' template calls for customization to specific datasets and models to optimize performance, given that certain templates may align better with particular dataset labels. Additionally, prompt ensembling (PT$_{\text{Ensemble}}$) outperformed individual prompts like CLS and PT$_{\text{Baseline}}$, but did not exceed PT$_{\text{Best}}$, which can be attributed to less effective prompts in the ensemble, potentially diminishing its overall efficacy.

\begin{table}[ht]
\centering
\resizebox{\columnwidth}{!}{
\begin{tabular}{ccc}
\toprule
\multirow{2}{*}{Dataset} & \multicolumn{2}{c}{Models} \\ \cline{2-3}
                         & LA & MS \\ \hline
{ESC50}      & \textit{Listen to}& \textit{A recording of}  \\ 
{US8K}       & \textit{I can hear} & \textit{Listen to an audio of} \\ 
{TUT2017}    & \textit{This is a sound track of} & \textit{Listen to an audio recording of} \\ 
{DCASE17-T4}  & \textit{A sound clip of} & \textit{This is a sound of}\\ 
{FSD50K}     & \textit{A sound recording of} & \textit{This is} \\ 
{AudioSet}   & \textit{This is an audio clip of} & \textit{This is a sound of} \\ 
\bottomrule

\end{tabular}
}
\caption{Best-performing prompt templates per dataset.}
\label{tab:best_prompt_templates}
\end{table}

\vspace{-2.em}
\subsection{Impact of class-specific descriptions}
In Table \ref{tab:results_extensive}, middle-panel, we assess the impact of class-specific descriptions on classification performance through our adaptive selection strategy, which determines which classes benefit from explicit descriptions. Our findings indicate that introducing class descriptions is indeed beneficial for disambiguating difficult classes, with audio-centric descriptions generally outperforming dictionary definitions. Focusing on model LA, class descriptions with contextual information (CD$_{\text{Context}}$) yielded the best results on average. While model MS also benefited from class-specific descriptions, it showed modest gains across datasets, likely due to its pretraining on a larger volume of data, including more audio-caption pairs.  For model MS, base audio-centric descriptions of class labels CD$_{\text{Base}}$ were the most effective, but still could not outperform prompt template-based methods in the top-panel for datasets such as DCASE17-T4, FSD50k and AudioSet. However, our adaptive strategy incorporating all types of descriptions (CD$_{\text{All}}$) did not generalize as effectively compared to individual setups, which was somewhat disappointing.

A comparison with state-of-the-art zero-shot audio classification scores reported in the literature, as shown in bottom line of Table \ref{tab:results_extensive}, reveals that our approach outperforms these benchmarks, including those utilizing prompt tuning strategies such as \cite{deshmukh2024domain}, across all evaluated datasets. The improvements are particularly notable for the US8K, TUT2017, FSD50K, and AudioSet datasets.

\vspace{-1.em}
\subsection{Disambiguation of classes through descriptions}
In Table \ref{tab:clas_improvements}, we show the top-3 classes with the greatest absolute improvement in classification using base descriptions compared to the simple use of class labels for the AudioSet and FSD50K datasets. We observe some words are ambiguous in meaning, for which an explicit description is beneficial as indicated by the large absolute improvements. A full list of relative improvements for all datasets is available on our companion website.
\vspace{-1.em}
\begin{table}[ht]
\centering
\resizebox{\columnwidth}{!}{
\begin{tabular}{ccc}
\toprule
Dataset & Class label & $\Delta$ Improvement [\%] \\ 
\hline
\multirow{3}{*}{AudioSet} & \textit{Bagpipes} & $+40.12$ \\
                          & \textit{Fire engine, fire truck (siren)} & $+39.79$ \\
                          & \textit{Gargling} & $+36.34$ \\
\hline
\multirow{3}{*}{FSD50K}   & \textit{Fowl} & $+67.75$ \\
                          & \textit{Scratching (performance technique)} & $+67.21$ \\
                          & \textit{Purr} & $+60.49$ \\
\bottomrule
\end{tabular}
}
\caption{Top-3 classes with highest absolute improved classification for model MS on AudioSet and FSD50K datasets using base audio-centric descriptions.}
\label{tab:clas_improvements}
\end{table}

\vspace{-2.em}
\section{Conclusion}
\label{sec:conclusion}
We demonstrated that prompt templates and class-specific descriptions can significantly impact the performance of zero-shot audio classification. While simple class labels can be highly effective, carefully crafted prompt templates and context-aware descriptions offer substantial improvements. Our findings advocate for a nuanced approach to prompt engineering, where the choice of format, content, and contextual information are tailored to the specific requirements of the model and dataset. Future work could explore automated methods for generating optimal prompts and descriptions, to further boost zero-shot audio classification scores.

\bibliographystyle{IEEEtran}
\bibliography{refs_formatted}

\begin{thebibliography}{10}
\providecommand{\url}[1]{#1}
\def\UrlFont{\rmfamily}
\providecommand{\newblock}{\relax}
\providecommand{\bibinfo}[2]{#2}
\providecommand\BIBentrySTDinterwordspacing{\spaceskip=0pt\relax}
\providecommand\BIBentryALTinterwordstretchfactor{4}
\providecommand\BIBentryALTinterwordspacing{\spaceskip=\fontdimen2\font plus
\BIBentryALTinterwordstretchfactor\fontdimen3\font minus
  \fontdimen4\font\relax}
\providecommand\BIBforeignlanguage[2]{{%
\expandafter\ifx\csname l@#1\endcsname\relax
\typeout{** WARNING: IEEEtran.bst: No hyphenation pattern has been}%
\typeout{** loaded for the language `#1'. Using the pattern for}%
\typeout{** the default language instead.}%
\else
\language=\csname l@#1\endcsname
\fi
#2}}

\bibitem{radford2021learning}
A.~Radford, J.~W. Kim, C.~Hallacy, A.~Ramesh, G.~Goh, S.~Agarwal, G.~Sastry,
  A.~Askell, P.~Mishkin, J.~Clark, \emph{et~al.}, ``Learning transferable
  visual models from natural language supervision,'' in \emph{Proc.
  ICML}.\hskip 1em plus 0.5em minus 0.4em\relax PMLR, 2021, pp. 8748--8763.

\bibitem{guzhov2022audioclip}
A.~Guzhov, F.~Raue, J.~Hees, and A.~Dengel, ``Audioclip: Extending clip to
  image, text and audio,'' in \emph{Proc. ICASSP}.\hskip 1em plus 0.5em minus
  0.4em\relax IEEE, 2022, pp. 976--980.

\bibitem{elizalde2024natural}
B.~Elizalde, S.~Deshmukh, and H.~Wang, ``Natural language supervision for
  general-purpose audio representations,'' in \emph{Proc. ICASSP}.\hskip 1em
  plus 0.5em minus 0.4em\relax IEEE, 2024, pp. 336--340.

\bibitem{wu2023large}
Y.~Wu, K.~Chen, T.~Zhang, Y.~Hui, T.~Berg-Kirkpatrick, and S.~Dubnov,
  ``Large-scale contrastive language-audio pretraining with feature fusion and
  keyword-to-caption augmentation,'' in \emph{Proc. ICASSP}.\hskip 1em plus
  0.5em minus 0.4em\relax IEEE, 2023, pp. 1--5.

\bibitem{manco2022contrastive}
I.~Manco, E.~Benetos, E.~Quinton, and G.~Fazekas, ``Contrastive audio-language
  learning for music,'' \emph{arXiv preprint arXiv:2208.12208}, 2022.

\bibitem{mei2023wavcaps}
X.~Mei, C.~Meng, H.~Liu, Q.~Kong, T.~Ko, C.~Zhao, M.~D. Plumbley, Y.~Zou, and
  W.~Wang, ``Wavcaps: A chatgpt-assisted weakly-labelled audio captioning
  dataset for audio-language multimodal research,'' \emph{arXiv preprint
  arXiv:2303.17395}, 2023.

\bibitem{liu2023audioldm}
H.~Liu, Z.~Chen, Y.~Yuan, X.~Mei, X.~Liu, D.~Mandic, W.~Wang, and M.~D.
  Plumbley, ``Audioldm: Text-to-audio generation with latent diffusion
  models,'' \emph{arXiv preprint arXiv:2301.12503}, 2023.

\bibitem{yuksekgonul2023and}
M.~Yuksekgonul, F.~Bianchi, P.~Kalluri, D.~Jurafsky, and J.~Zou, ``When and why
  vision-language models behave like bags-of-words, and what to do about it?''
  in \emph{Proc. ICLR}, 2023.

\bibitem{anperceptionclip}
B.~An, S.~Zhu, M.-A. Panaitescu-Liess, C.~K. Mummadi, and F.~Huang,
  ``Perceptionclip: Visual classification by inferring and conditioning on
  contexts,'' in \emph{Proc. ICLR}, 2024.

\bibitem{salinas2024butterfly}
A.~Salinas and F.~Morstatter, ``The butterfly effect of altering prompts: How
  small changes and jailbreaks affect large language model performance,''
  \emph{arXiv preprint arXiv:2401.03729}, 2024.

\bibitem{pratt2023does}
S.~Pratt, I.~Covert, R.~Liu, and A.~Farhadi, ``What does a platypus look like?
  generating customized prompts for zero-shot image classification,'' in
  \emph{Proc. ICCV}, 2023, pp. 15\,691--15\,701.

\bibitem{roth2023waffling}
K.~Roth, J.~M. Kim, A.~Koepke, O.~Vinyals, C.~Schmid, and Z.~Akata, ``Waffling
  around for performance: Visual classification with random words and broad
  concepts,'' in \emph{In Proc. of the IEEE/CVF International Conference on
  Computer Vision}, 2023, pp. 15\,746--15\,757.

\bibitem{mirza2024lafter}
M.~J. Mirza, L.~Karlinsky, W.~Lin, H.~Possegger, M.~Kozinski, R.~Feris, and
  H.~Bischof, ``Lafter: Label-free tuning of zero-shot classifier using
  language and unlabeled image collections,'' \emph{Advances in Neural
  Information Processing Systems}, vol.~36, 2024.

\bibitem{elizalde2023clap}
B.~Elizalde, S.~Deshmukh, M.~Al~Ismail, and H.~Wang, ``Clap learning audio
  concepts from natural language supervision,'' in \emph{Proc. ICASSP}.\hskip
  1em plus 0.5em minus 0.4em\relax IEEE, 2023, pp. 1--5.

\bibitem{kushwahamultimodal}
S.~S. Kushwaha and M.~Fuentes, ``A multimodal prototypical approach for
  unsupervised sound classification.''

\bibitem{li2024audio}
Y.~Li, X.~Wang, and H.~Liu, ``Audio-free prompt tuning for language-audio
  models,'' in \emph{Proc. ICASSP}.\hskip 1em plus 0.5em minus 0.4em\relax
  IEEE, 2024, pp. 491--495.

\bibitem{deshmukh2024domain}
S.~Deshmukh, R.~Singh, and B.~Raj, ``Domain adaptation for contrastive
  audio-language models,'' \emph{arXiv e-prints}, pp. arXiv--2402, 2024.

\bibitem{jiang2023mistral}
A.~Q. Jiang, A.~Sablayrolles, A.~Mensch, C.~Bamford, D.~S. Chaplot, D.~d.~l.
  Casas, F.~Bressand, G.~Lengyel, G.~Lample, L.~Saulnier, \emph{et~al.},
  ``Mistral 7b,'' \emph{arXiv preprint arXiv:2310.06825}, 2023.

\bibitem{oncescu2024sound}
A.-M. Oncescu, J.~F. Henriques, A.~Zisserman, S.~Albanie, and A.~S. Koepke, ``A
  sound approach: Using large language models to generate audio descriptions
  for egocentric text-audio retrieval,'' in \emph{Proc. ICASSP}.\hskip 1em plus
  0.5em minus 0.4em\relax IEEE, 2024, pp. 7300--7304.

\bibitem{liu2019roberta}
Y.~Liu, M.~Ott, N.~Goyal, J.~Du, M.~Joshi, D.~Chen, O.~Levy, M.~Lewis,
  L.~Zettlemoyer, and V.~Stoyanov, ``Roberta: A robustly optimized bert
  pretraining approach,'' \emph{arXiv preprint arXiv:1907.11692}, 2019.

\bibitem{radford2019language}
A.~Radford, J.~Wu, R.~Child, D.~Luan, D.~Amodei, I.~Sutskever, \emph{et~al.},
  ``Language models are unsupervised multitask learners,'' \emph{OpenAI blog},
  vol.~1, no.~8, p.~9, 2019.

\bibitem{chen2022hts}
K.~Chen, X.~Du, B.~Zhu, Z.~Ma, T.~Berg-Kirkpatrick, and S.~Dubnov, ``Hts-at: A
  hierarchical token-semantic audio transformer for sound classification and
  detection,'' in \emph{Proc. ICASSP}.\hskip 1em plus 0.5em minus 0.4em\relax
  IEEE, 2022, pp. 646--650.

\bibitem{piczak2015dataset}
\BIBentryALTinterwordspacing
K.~J. Piczak, ``{ESC}: {Dataset} for {Environmental Sound Classification},'' in
  \emph{Proc. ACM-MM}.\hskip 1em plus 0.5em minus 0.4em\relax {ACM Press}, pp.
  1015--1018. [Online]. Available:
  \url{http://dl.acm.org/citation.cfm?doid=2733373.2806390}
\BIBentrySTDinterwordspacing

\bibitem{salamon2014dataset}
J.~Salamon, C.~Jacoby, and J.~P. Bello, ``A dataset and taxonomy for urban
  sound research,'' in \emph{Proc. ACM-MM}, 2014, pp. 1041--1044.

\bibitem{mesaros2019sound}
A.~Mesaros, A.~Diment, B.~Elizalde, T.~Heittola, E.~Vincent, B.~Raj, and
  T.~Virtanen, ``Sound event detection in the dcase 2017 challenge,''
  \emph{Proc. IEEE/ACM Trans. Audio Speech Lang.}, vol.~27, no.~6, pp.
  992--1006, 2019.

\bibitem{fonseca2021fsd50k}
E.~Fonseca, X.~Favory, J.~Pons, F.~Font, and X.~Serra, ``Fsd50k: an open
  dataset of human-labeled sound events,'' \emph{IEEE/ACM Trans. Audio Speech
  Lang.}, vol.~30, pp. 829--852, 2021.

\bibitem{gemmeke2017audio}
J.~F. Gemmeke, D.~P. Ellis, D.~Freedman, A.~Jansen, W.~Lawrence, R.~C. Moore,
  M.~Plakal, and M.~Ritter, ``Audio set: An ontology and human-labeled dataset
  for audio events,'' in \emph{Proc. ICASSP}.\hskip 1em plus 0.5em minus
  0.4em\relax IEEE, 2017, pp. 776--780.

\bibitem{sclarquantifying}
M.~Sclar, Y.~Choi, Y.~Tsvetkov, and A.~Suhr, ``Quantifying language models'
  sensitivity to spurious features in prompt design or: How i learned to start
  worrying about prompt formatting,'' in \emph{Proc. ICLR}, 2024.

\end{thebibliography}

%
%
%
%
%
%
%
%
%

\end{sloppy}
\end{document}